\documentclass[%
aip,
 amsmath,amssymb,
 reprint,%
]{revtex4-1}

\usepackage{graphicx}
\usepackage{dcolumn}
\usepackage{bm}

\usepackage[utf8]{inputenc}
\usepackage[T1]{fontenc}
\usepackage{mathptmx}

\usepackage{siunitx}
\usepackage{hyperref}

\hypersetup{bookmarksnumbered=true,
  colorlinks = true, 	
  allcolors = blue,		
  breaklinks = true,	
  backref = true		
}
\begin{document}

\preprint{AIP/123-QED}

\title[]{Fast low-noise transimpedance amplifier for scanning tunneling microscopy and beyond}

\author{Martin \v{S}tubian}
 \altaffiliation[Also at ]{Brno University of Technology, 60190 Brno, Czech Republic}
 \affiliation{Institute of Applied Physics, TU Wien, 1040 Vienna, Austria}

\author{Juraj Bobek}
 \altaffiliation[Also at ]{Brno University of Technology, 60190 Brno, Czech Republic}
 \affiliation{Institute of Applied Physics, TU Wien, 1040 Vienna, Austria}

\author{Martin Setvin}%
\affiliation{Institute of Applied Physics, TU Wien, 1040 Vienna, Austria}
 \affiliation{Department of Surface and Plasma Science, Faculty of Mathematics and Physics, Charles University, 180 00 Prague 8, Czech Republic}

\author{Ulrike Diebold}%
\affiliation{Institute of Applied Physics, TU Wien, 1040 Vienna, Austria}

\author{Michael Schmid}%
 \email{schmid@iap.tuwien.ac.at}
\affiliation{Institute of Applied Physics, TU Wien, 1040 Vienna, Austria}

\date{2 June 2020}

\begin{abstract}
Accepted manuscript, published in Rev. Sci. Instrum. 91, 074701 (2020);
DOI: \href{https://doi.org/10.1063/5.0011097}{10.1063/5.0011097}
\\[12pt]
A transimpedance amplifier has been designed for scanning tunneling microscopy (STM). The amplifier features low noise (limited by the Johnson noise of the 1\,\si{\giga\ohm} feedback resistor at low input current and low frequencies), sufficient bandwidth for most STM applications (50\,kHz at 35\,pF input capacitance), a large dynamic range ($0.1$\,pA to 50\,nA without range switching) as well as a low input voltage offset. The amplifier is also suited for placing its first stage into the cryostat of a low-temperature STM, minimizing the input capacitance and reducing the Johnson noise of the feedback resistor. The amplifier may also find applications for specimen current imaging and electron-beam induced current measurements in scanning electron microscopy, and as a photodiode amplifier with a large dynamic range. 
This paper also discusses sources of noise including the often neglected effect of non-balanced input impedance of operational amplifiers, and describes how to accurately measure and adjust the frequency response of low-current transimpedance amplifiers.
\end{abstract}

\maketitle

\section{\label{sec:introduction}Introduction}

Transimpedance amplifiers, also known as current-to-voltage converters (I/V converters) have many applications, for photodiode signals, \cite{sackinger_2017} biophysics, scanning tunneling microscopy (STM), and much more. Due to parasitic capacitances, there is a tradeoff between bandwidth (speed) on the one hand and sensitivity (noise) on the other. For STM applications, the following properties are essential:\\
- Low noise and high sensitivity. STM is not an innocent probe; even currents in the pA range can modify surfaces. \cite{scheiber_2010} Imaging poorly conducting samples such as wide-bandgap materials requires working at low currents. Thus, the amplifier should provide good performance at least down to the low pA range, preferably below 1\,pA.\\
- Large dynamic range. Applications like single-atom manipulation (tip-sample resistance $R_\mathrm{t} \approx 50$--$500$\,\si{\kilo\ohm})\cite{eigler_positioning_1990,bartels_1997} and the measurement of conductive channels between the sample and the tip upon contact formation (tip-sample resistance $R_\mathrm{t} \approx 10$\,\si{\kilo\ohm})\cite{gimzewski_1987} require high currents of at least tens of nA.
STM imaging of metals with chemical contrast (tip-sample resistance $R_\mathrm{t} \approx 100$\,\si{\kilo\ohm}--1\,\si{\mega\ohm})\cite{schmid_direct_1993,hebenstreit_pt25rh75111_1999} is often done at tunneling currents around 5--10\,nA. In constant-current mode, for adequate response of the feedback loop, the current range should be larger than the average tunneling current (i.e., the current setpoint) by more than a factor of two. When also considering the requirements for low-current imaging discussed above, a dynamic range in the order of $10^5$ is desirable for a general-purpose STM preamplifier.  \\
- Sufficient bandwidth. The bandwidth of the amplifier limits the speed of data acquisition. While video-rate STM \cite{rost_scanning_2005} requires a bandwidth in the 500\,kHz--1\,MHz range, most STM controllers cannot handle such high data rates and are limited to sampling rates of a few 10\,kHz. Equally important, in constant-current mode of the STM, the preamplifier should not introduce substantial phase shifts in the frequency range where the feedback loop is active, since this will reduce the stability (phase margin) of the feedback.\\
- Low input offset voltage. Measurements at low tip-sample resistance require working at low bias voltages ($\approx 1$\,mV). High-resolution scanning tunneling spectroscopy at cryogenic temperatures requires sub-mV voltage accuracy. \\
- Low input impedance. The input impedance of the amplifier should be well below the tip-sample resistance $R_\mathrm{t}$. For a combination of STM with non-contact atomic force microscopy (ncAFM), a low input impedance at the resonance frequency of the ncAFM sensor (typically $\approx 30$\,kHz) is also required to keep the bias voltage between tip and sample constant, otherwise the voltage modulation by the variations of the input current causes electrostatic forces that excite the ncAFM sensor \cite{majzik_simultaneous_2012}.\\
- No phase inversion. Voltage pulses, mobile species at the surface, or loose flakes (e.g. of a graphite sample) can lead to a sudden short of the tunneling junction, overloading the preamplifier. Some operational amplifiers produce output voltages with the wrong polarity under overload conditions (``phase inversion''). When detecting a tunneling current with the wrong polarity most STM controllers would crash the tip into the sample. Therefore, one should select operational amplifiers not susceptible to phase inversion.\\
- Finally, especially for low-temperature STMs with a long cable between the STM (housed in a cryostat) and atmospheric-side electronics, the possibility of having the first amplifier stage in vacuum is desirable. As will be discussed below, the capacitance of a long cable is detrimental not only for the bandwidth but also for the noise performance of the amplifier.

\begin{figure}
\includegraphics[width=8cm]{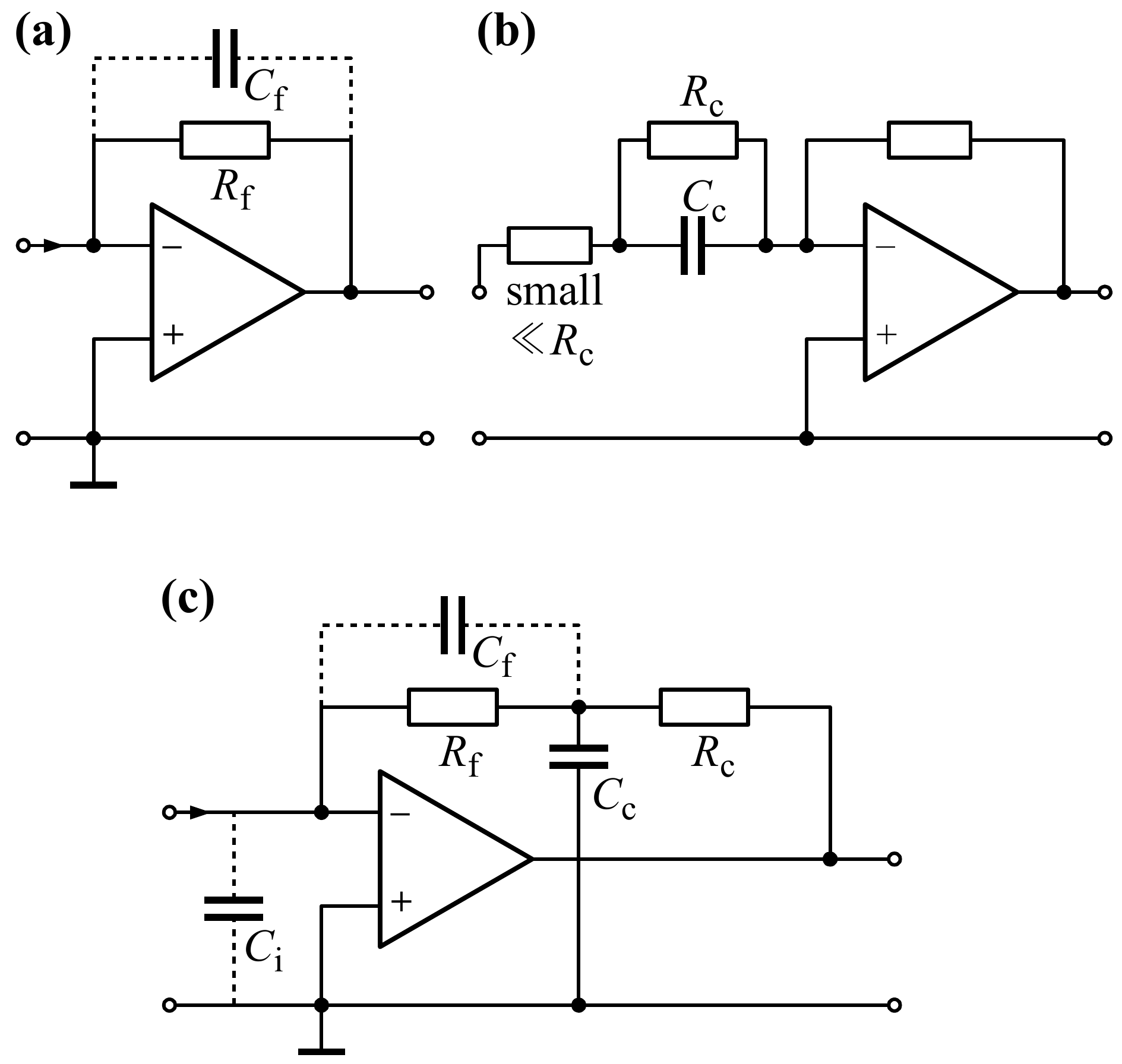}
\caption{\label{fig1_basics} (a) Basic design of a transimpedance amplifier, including the parasitic capacitance parallel to the feedback resistor, and (b) a second stage for compensating the resulting low-pass behavior. (c) Basic design of a transimpedance amplifier with compensation in the feedback loop. \cite{michel_low-temperature_1992,giusi_ultra-low-noise_2015}}
\end{figure}

Figure \ref{fig1_basics}(a) shows the basic design of a transimpedance amplifier. Due to the unavoidable parasitic capacitance $C_\mathrm{f}$ of the feedback resistor $R_\mathrm{f}$, the bandwidth of this circuit is limited to $\omega_{-3\,\mathrm{dB}}=1/(R_\mathrm{f}C_\mathrm{f})$. For a feedback resistor of 1\,\si{\giga\ohm} and a typical parasitic capacitance of 0.1\,pF, this corresponds to $f_{-3\,\mathrm{dB}}=1.6$\,kHz, much less than the desired bandwidth. Placing several resistors in series (each with lower resistance) can lead to some improvement, \cite{ciofi_how_2006} but not reach the desired bandwidth of at least tens of kHz. The easiest solution for the bandwidth problem is using a lower feedback resistor, either driven by a voltage divider \cite{lin_2012} or with post-amplification. As discussed below, this comes at the cost of increased (Johnson) noise.

The low-pass behavior caused by the stray capacitance $C_\mathrm{f}$ can be compensated by a second stage having a gain increasing at high frequencies, as shown in Fig.\ \ref{fig1_basics}(b). \cite{carla_development_2004,paul_updated_2006,ciofi_new_2007,ferrari_ultra-low-noise_2009} It has been noted earlier \cite{michel_low-temperature_1992} that then the input voltage noise of the second stage becomes an issue at high frequencies (where its gain must be high to compensate for the low-pass behavior of the first stage). Although the design of the amplifier in Ref.\ \onlinecite{rost_scanning_2005} has not been published, it is likely that it involves a similar circuit and the increase of its noise being more than proportional to the bandwidth \cite{rost_scanning_2005} is due to this problem.

Another method of compensating for the low-pass behavior of the amplifier in Fig.\ \ref{fig1_basics}(a) is adding a low-pass filter to the feedback path, again with $R_\mathrm{c}C_\mathrm{c} = R_\mathrm{f}C_\mathrm{f}$ [Fig.\ \ref{fig1_basics}(c)].\cite{michel_low-temperature_1992,giusi_ultra-low-noise_2015}  The basic circuit in Fig.\ \ref{fig1_basics}(c) is not stable, however: At high frequencies, $\omega \gg 1/(R_\mathrm{f}C_\mathrm{f})$, where $R_\mathrm{f}$ plays no role, $C_\mathrm{f}$ forms a capacitive voltage divider with the parasitic input capacitance $C_\mathrm{i}$ (including the input capacitance of the operational amplifier, and the capacitance of the cable from the input to the amplifier, if any) with a phase shift near \ang{0}. Thus, the phase shift of the $R_\mathrm{c}C_\mathrm{c}$ low-pass circuit will add to the phase shift of the operational amplifier's open loop gain; both near \ang{-90} at high frequencies, leading to a phase margin close to \ang{0}. This problem can be solved by either a capacitor parallel to $R_\mathrm{c}$ \cite{michel_low-temperature_1992} or using an amplifier with a low phase shift in the relevant frequency range, i.e., an amplifier with fixed (and rather limited) gain. \cite{giusi_ultra-low-noise_2015} The first approach, though delivering excellent performance, is not trivial since it either involves a capacitive load to the operational amplifier (which may also induce instability) or a high $R_\mathrm{c}$ value, which makes the node between $R_\mathrm{f}$ and $R_\mathrm{c}$ sensitive to pickup of interference signals. This approach also requires a very careful design of the environment of the feedback resistor to ensure stability of the circuit in spite of various stray capacitances. Furthermore, it is unclear in which range of input capacitances the circuit in  Ref.\ \onlinecite{michel_low-temperature_1992} can be stable if there is a cable between the STM and the amplifier (only cables up to 2\,cm length are mentioned in Ref.\ \onlinecite{michel_low-temperature_1992}). The performance of this circuit is impressive, however, with a bandwidth of 1\,MHz reported even with a feedback resistor of 10\,\si{\giga\ohm} \cite{michel_low-temperature_1992}. The main drawback of the circuit in Ref.\ \onlinecite{michel_low-temperature_1992} is the use of a dual-JFET input with a large input offset voltage ($\approx 25$\,mV), not acceptable for many STM applications.

The other approach to obtain stable operation of the circuit in Fig.\ \ref{fig1_basics}(c), using an amplifier with fixed gain (and low phase shift) \cite{giusi_ultra-low-noise_2015}, is a known solution of the $C_\mathrm{i}$-induced stability problem \cite{smith_wide-band_1997}. The operational amplifier in Fig.\ \ref{fig1_basics}(c) was replaced by a circuit consisting of a noninverting input stage followed by an inverting stage (in Ref.\ \onlinecite{giusi_ultra-low-noise_2015}, with gains of 25.9 and $-500$, respectively). The overall bandwidth was reported to be around 100\,kHz with $R_\mathrm{f} = 1$\,\si{\giga\ohm}, clearly sufficient for most STM applications, and the circuit also works at input capacitance values of 47 or 100\,pF, though with reduced bandwidth. For STM applications, it has to be noted, however, that the input impedance of this amplifier is roughly equal to $R_\mathrm{f}/|A|$, where $A$ is the gain of the amplifier. With the values of Ref.\ \onlinecite{giusi_ultra-low-noise_2015}, the DC input impedance is 77\,\si{\kilo\ohm}, so switching of the feedback resistor would be required for measurements involving low values of the tip-sample resistance $R_\mathrm{t}$. A further problem comes from the fact that the input of this amplifier is at the noninverting input of the operational amplifier, which is on the IC housing usually next to the negative supply. Thus, a leakage resistance between neighboring pins of 1\,\si{\tera\ohm} will cause a leakage current of a several pA (depending on the supply voltage).
\footnote{We are aware of only one operational amplifier where the noninverting input is not neighboring a supply pin, the femtoamp-sensitive ADA4530. Its input voltage noise and low speed make it less attractive for an STM preamplifier except when working at extremely low currents and keeping the input capacitance very low. For some operational amplifiers, such as the OPA627/OPA637, a similar problem arises for the the inverting input due to a neighboring offset trim pin; this is not the case for the AD8615 and OPA657.}

The amplifier described in the current paper is based on the circuit of Figure \ref{fig1_basics}(c), but solves the stability problem by placing a resistor in series to $C_\mathrm{c}$. Before discussing the details of the circuit and its performance, the following two sections will be devoted to basics of the noise of transimpedance amplifiers and the measurement of their frequency response.

\section{\label{sec:noise}Noise Considerations}

\begin{figure}
\includegraphics[width=8cm]{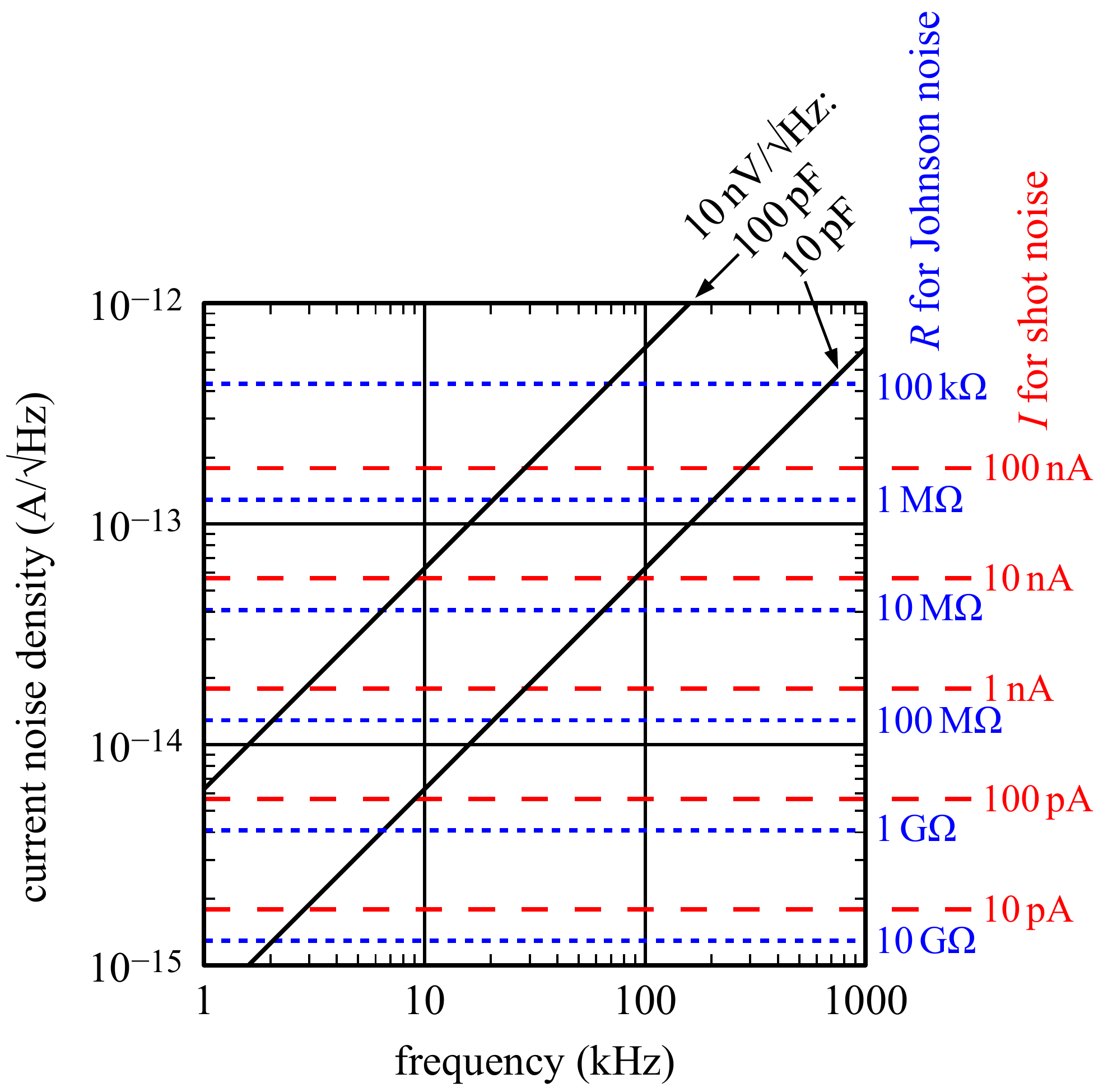}
\caption{\label{fig2_noise} Contributions to the noise of a transimpedance amplifier. The thermal (Johnson) noise for different values of the feedback resistor at $T = 300$\,K is given by blue, short dashes, the shot noise for different current values by red, long dashes. The diagonal lines show the impact of the voltage noise of the operational amplifier assuming a frequency-independent input voltage noise of 10\,nV/$\sqrt\mathrm{Hz}$ and (effective) input capacitance values of 10 and 100\,pF.}
\end{figure}

For reducing the noise, several contributions have to be taken into account (Fig.\ \ref{fig2_noise}). The thermal (Johnson) current noise density of the feedback resistor, $\sqrt{4 k_\mathrm{B} T/R_\mathrm{f}}$, decreases with increasing value of $R_\mathrm{f}$. As described above, a high value of $R_\mathrm{f}$ leads to a lower bandwidth, furthermore the input current range also decreases. Thus, there is no point in increasing $R_\mathrm{f}$ if other sources of noise dominate. Except for low input currents $I$, the shot noise, $\sqrt{2 e |I|}$ (with $e$ being the elementary charge), has to be considered. It follows from these equations that the shot noise dominates over the thermal noise if the voltage drop across the feedback resistor is higher than $V=2 k_\mathrm{B} T/e$, which is 52\,mV for the feedback resistor at 300\,K. Thus, assuming that the input current range is limited by the maximum voltage at the feedback resistor of $\pm 10$\,V (provided by a typical opamp output), the Johnson noise will dominate only if the input current is 0.5\% of the full scale or less. Considering the large dynamic range desirable for an STM current amplifier, it is nevertheless important to reduce the Johnson noise (keeping $R_\mathrm{f}$ as high as possible, and taking advantage of keeping it cold in low-temperature STMs) for optimum performance at low currents.

The input current noise of operational amplifiers relevant for this application is typically $\approx 1$\,fA$\sqrt{\mathrm{Hz}}$ or better, below the thermal noise of a 1\,\si{\giga\ohm} resistor (4.1\,fA$/\sqrt{\mathrm{Hz}}$ at room temperature)
\footnote{The input current noise of the AD8615 is given as 0.05\,pA$/\sqrt\mathrm{Hz}$ at 1\,kHz in its datasheet. This value is probably a printing mistake. At 1\,kHz, our amplifiers using the AD8615 show a noise limited by the Johnson noise of the 1\,\si{\giga\ohm} feedback resistor, indicating that the input current noise of the AD8615 is below 2\,fA$/\sqrt\mathrm{Hz}$ at room temperature. This is also in line with the usual expectation that the current noise is not much higher than the shot noise of the input bias current (0.25\,fA$/\sqrt\mathrm{Hz}$ for a typical bias current of 0.2\,pA) or the shot noise of the leakage current of the input protection diodes, whichever is larger. Assuming that the input protection diodes to the negative and positive supply are well matched and their leakage currents differ by only 10\%, and this difference causes the input bias current, they would leak $10\times$ the input bias current (2\,pA), resulting in a shot noise of roughly 1\,fA$/\sqrt\mathrm{Hz}$}.
The input voltage noise $v_\mathrm{n}$, together with the input capacitance $C_\mathrm{i}$ (of the amplifier, plus stray capacitance and that of the cable, if any), is usually more important. Assuming an otherwise ideal operational amplifier and feedback network, the current through the input capacitor caused by the input voltage noise $v_\mathrm{n}$ is supplied by the feedback resistor, and thus measured the same way as the input current
\footnote{Here we assume that the real part of the input impedance is high enough so that the current created in it by $v_\mathrm{n}$ can be neglected. This is usually the case for the resistance $R_\mathrm{t}$  of the tunneling junction in an STM at low and moderate currents. At high currents the shot noise dominates anyhow. The same reasoning usually holds for photodiode amplifiers}.
This noise contribution therefore corresponds to an input current noise of
\begin{equation}
  i_\mathrm{n} = \omega C_\mathrm{i} v_\mathrm{n}\ .
    \label{eq:noiseCin}
\end{equation}
For amplifiers with large bandwidth, this source of noise will dominate (Fig.\ \ref{fig2_noise}), which underlines the importance of keeping the input capacitance as low as possible. This can be easily explained by the operational amplifier being a voltage amplifier: The input current needs to be converted to a voltage to be detected; if the input is bypassed by a capacitor, that voltage will be lower and more susceptible to the influence of voltage noise.

\begin{figure}
\includegraphics[width=6cm]{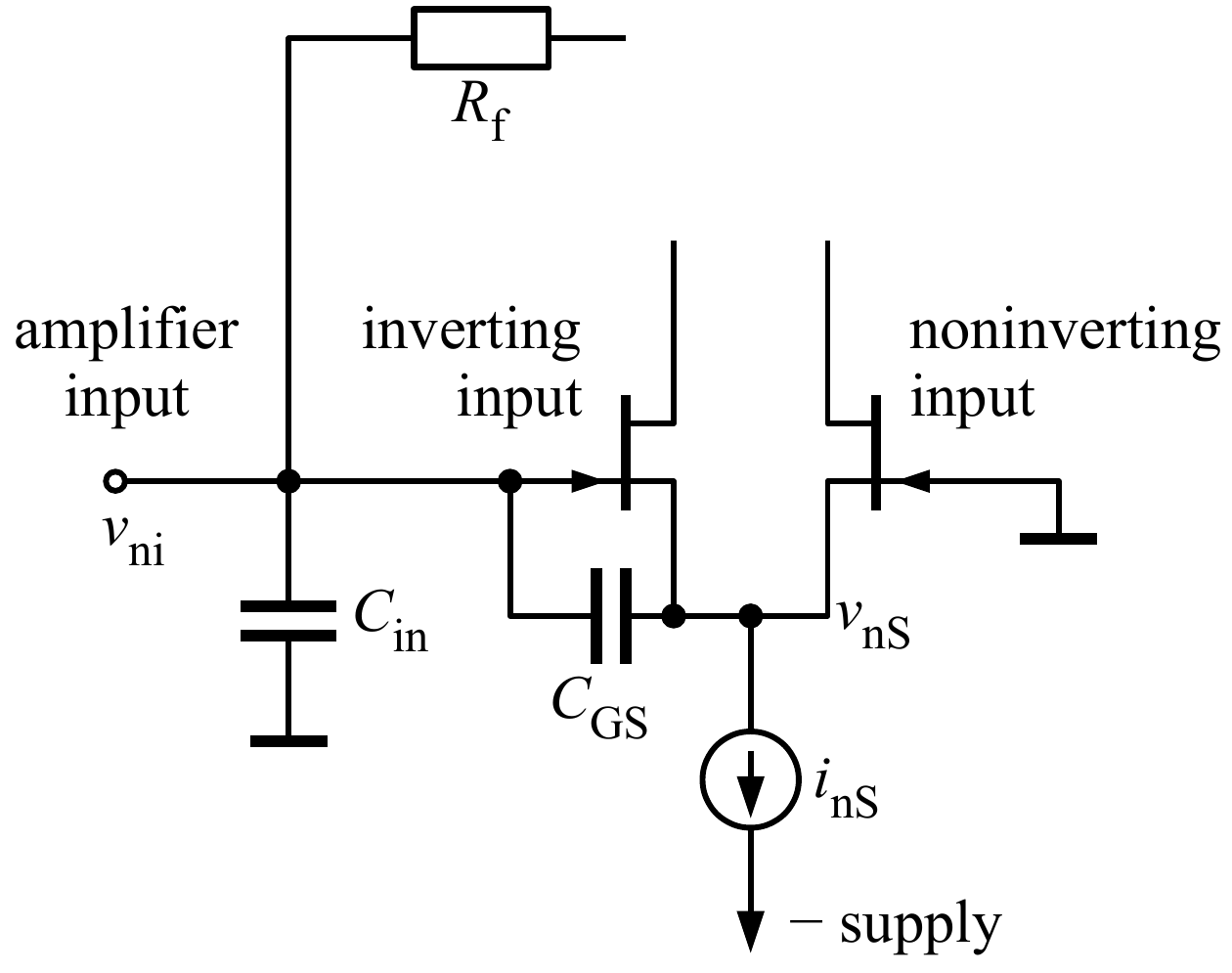}
\caption{\label{fig3_inputNoise} Simplified view of the input stage of an operational amplifier used as transimpedance amplifier. Noise of the current source in the input stage leads to a voltage noise $v_\mathrm{nS}$ at the source terminals of the FETs. Due to the unbalanced input impedances, this leads to a noise voltage $v_\mathrm{ni}$ at the input. The circuit has been drawn for n-channel JFETs, but the same applies for p-channel FETs and MOSFETs in the input stage.}
\end{figure}%
Our experiments provide evidence of an additional noise contribution that also increases proportional to the frequency and is usually not considered. It has been noted previously that the experimentally observed noise of transimpedance amplifiers could be explained if the input capacitances of operational amplifiers were higher than stated in datasheets \cite{masalov_noise_2017}. We suggest the following reason for this: The input voltage noise of operational amplifiers is usually measured with a very low input impedance for both inputs. Transimpedance amplifiers have a high input impedance for the non-inverting input, however. Figure \ref{fig3_inputNoise} shows a simplified view of an operational amplifier's input stage in such a configuration. Assuming a noise contribution $i_\mathrm{nS}$ of the current supply for the source terminals of the input FETs, a noise voltage $v_\mathrm{nS}$ will appear there, and be coupled into the gate by the gate-source capacitance $C_\mathrm{GS}$. As the gate of the other FET is tied to ground, it acts as a differential-mode input voltage. This noise contribution is related to a noise source of field effect transistors named ``induced gate noise'' (IGN) \cite{jazaeri_modeling_2015}. This noise type has been found early and been linked to noise of the drain current, such as shot noise, leading to a modulation of the channel potential, which couples capacitively to the gate \cite{van_der_ziel_thermal_1962,van_der_ziel_gate_1963}. Therefore, IGN can be observed in common-source circuits. In our case, we consider it likely that the main source of the noise of the FET channel current does not have its root in the FETs themselves (the FETs do not determine the overall current) but rather in the current source driving the input FETs. In any case, even if IGN contributions in the usual sense play a role, they can be added to the noise $v_\mathrm{nS}$ and will affect the gate voltage through essentially the same (gate-source or gate-channel) capacitance $C_\mathrm{GS}$.
The noise at the common source point of the FETs, $v_\mathrm{nS}$, leads to a noise voltage at the inverting input (FET gate),
\begin{equation}
  v_\mathrm{ni} = v_\mathrm{nS} \frac{C_\mathrm{GS}}{C_\mathrm{in} + C_\mathrm{GS}}\ ,
    \label{eq:noiseUnbalanced1}
\end{equation}
where $C_\mathrm{in}$ is the capacitance between the input and ground (not exactly identical to the $C_\mathrm{i}$ mentioned above; $C_\mathrm{i}$ also includes contributions from $C_\mathrm{GS}$). Similar to the ``usual'' input voltage noise $v_\mathrm{n}$ of the operational amplifier, this noise voltage has to be compensated by a current through $R_\mathrm{f}$, which yields a noise current contribution of
\begin{equation}
  i_\mathrm{n} = \omega C_\mathrm{i} v_\mathrm{nS} \frac{C_\mathrm{GS}}{C_\mathrm{in} + C_\mathrm{GS}}
    \approx \omega v_\mathrm{nS} C_\mathrm{GS}\ .
      \label{eq:noiseUnbalanced2}
\end{equation}%
The latter approximation is justified if $C_\mathrm{GS} \ll C_\mathrm{i}$ or $C_\mathrm{i} \approx C_\mathrm{in} + C_\mathrm{GS}$, which is usually fulfilled. This contribution has the same frequency dependence as (\ref{eq:noiseCin}), thus the two can be added
\footnote{We can assume that these two sources of noise are uncorrelated, because (\ref{eq:noiseCin}) is noise of the input FET and (\ref{eq:noiseUnbalanced2}) is due to noise of the current source. Therefore the squares (noise powers) have to be added and the current noise is the square root of the sum.}
and also written as in (\ref{eq:noiseCin}), but with a larger, effective input capacitance
\begin{equation}
  C_\mathrm{eff} \approx  \sqrt{ C_\mathrm{i}^2+ \frac{v_\mathrm{nS}^2}{v_\mathrm{n}^2} C_\mathrm{GS}^2}\ .
      \label{eq:Ceff}
\end{equation}

 In a circuit with balanced input impedances, or in amplifiers where the input impedance of both inputs is low, the source of noise in (\ref{eq:noiseUnbalanced1}), (\ref{eq:noiseUnbalanced2}) can be neglected. For some operational amplifier types, we have indeed observed a reduction of noise in a balanced circuit, where the noninverting input is connected to ground via a resistor equal to $R_\mathrm{f}$ and a capacitor equal to the input capacitance. Unfortunately, for our application this approach turned out to be impractical, because (i) that resistor increases the noise at low frequencies (additional Johnson noise), (ii) it reduces the bandwidth of the circuit, and (iii) it introduces another adjustment point (the input capacitance and the capacitance in the noninverting branch have to be matched), and adjusting that balancing capacitance requires readjusting the frequency compensation. Thus we opt to select operational amplifier types where the ``additional'' high-frequency noise (beyond that expected from $v_\mathrm{n}$ and $C_\mathrm{i}$) is low. In principle, problem (i) could be alleviated by using a lower resistor value than $R_\mathrm{f}$ at the noninverting input (resulting lower voltage noise). The resulting imbalance will be relevant at low frequencies only, where the noise contribution of (\ref{eq:noiseUnbalanced2}) is negligible. First experiments in this direction did not yield a noise reduction as high as with a value of $R_\mathrm{f}$ at the noninverting input, however. Nevertheless, support for this analysis comes from observation that some operational amplifiers benefit more from a balanced design than others, and those that benefit more indeed have a large contribution of this ``additional'' high-frequency noise beyond that expected from eq.\ (\ref{eq:noiseCin}) (cf.\ section \ref{sec:performance}). A further observation consistent with our model is the increase of noise of the AD8615 when the positive supply voltage is lower than $\approx 2$\,V; we consider it likely that the current source of the p-channel input MOSFETs becomes more noisy when its headroom (voltage drop) is too low. Furthermore, we believe that the suppression of this type of noise in a balanced design is responsible for the exceptionally good performance of the preamplifier described in Ref.\ \onlinecite{huber_2017} for ncAFM sensors.

Another possible noise contribution is excess noise of the resistor (i.e., noise in addition to the Johnson and shot noise). This noise contribution is usually related to resistance fluctuations; unless low-quality resistors are used, it can be neglected at low currents, where the noise performance of a transimpedance amplifier is most critical.

\section{\label{sec:measurement}Measuring the Frequency Response}

Measuring the frequency response of a transimpedance amplifier is not simple: If the input current is derived from a frequency generator with a series resistor, its stray capacitance will strongly influence the result. It is difficult to determine that stray capacitance with sufficiently high accuracy to correct for it numerically. Similar as for the stray capacitance of the feedback resistor, this problem can be alleviated by using low resistance values (though at the cost of increased noise) and placing several of them in series \cite{yang_enhanced_2019}. We use a much simpler method, putting a small capacitor (1\,pF) between the frequency generator and the amplifier input. It is easy to correct for the capacitor's frequency-dependent impedance in the data analysis, and any stray capacitance will only affect the absolute value of the gain, not the frequency dependence or the phase. If required, the absolute value of the gain can be easily determined by a separate DC measurement. With the amplifier already connected to an STM, the frequency response can be measured in exactly the same fashion by retracting the tip and using an AC bias voltage and the tip-sample capacitance as a high-impedance AC current source.

\begin{figure*}
\includegraphics[width=17cm]{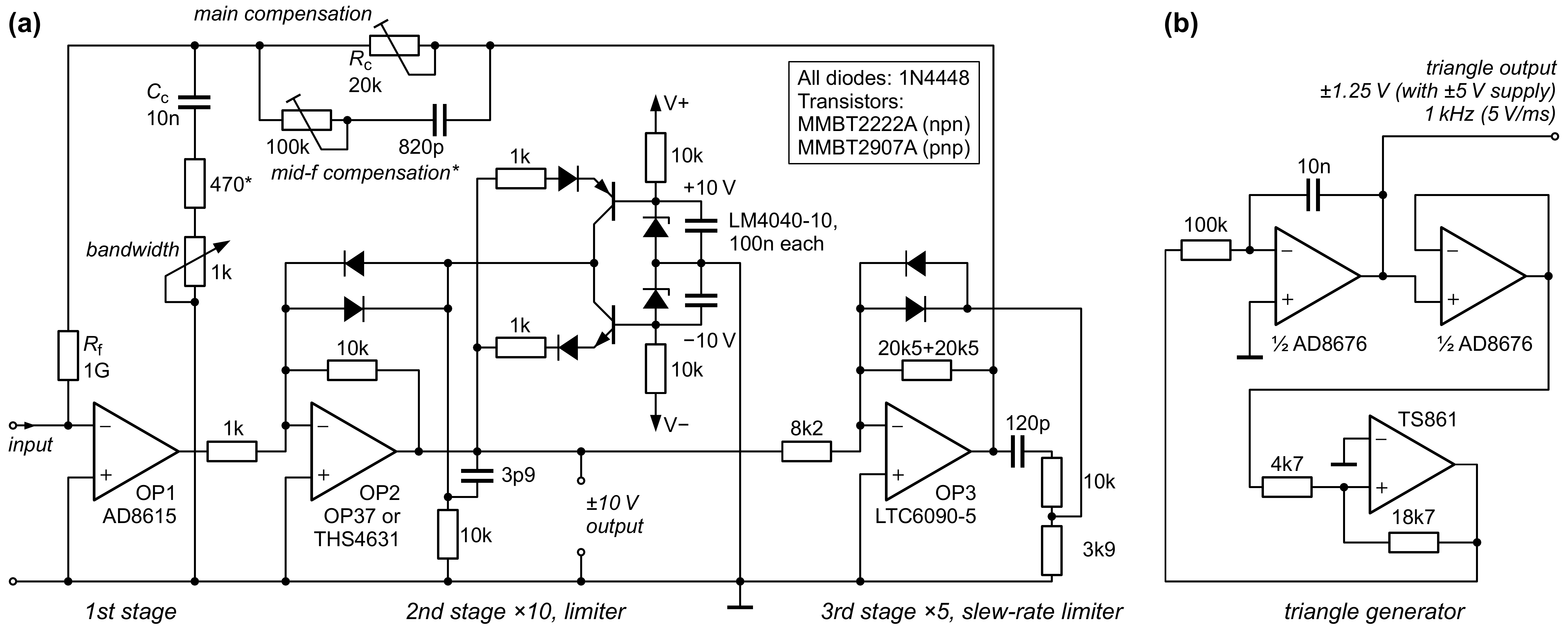}
\caption{\label{fig4_circuit} (a) Schematics of the transimpedance amplifier circuit. See the text for components marked with an asterisk. (b) The triangle generator for adjustment (see section \ref{sec:measurement}) is a simple integrator-comparator design with an additional buffer to reduce influencing the integrator by transients from the comparator.}
\end{figure*}

For an amplifier with compensation of parasitic capacitances, there will be usually adjustment points such as potentiometers to obtain a flat frequency response. Using a small capacitor at the input also provides an elegant way for adjusting: When applying a triangle waveform via a small capacitor to the input, the transimpedance amplifier will act as a differentiator, and, with a flat frequency response, provide a square wave at the output. Adjustment of the frequency response of the amplifier is then similar to adjusting a 1:10 oscilloscope probe. In our STM amplifier, we have added a 1\,kHz triangle generator [Fig.\ \ref{fig4_circuit}(b)], whose output can be used instead of the STM bias. With the tip not in tunneling contact, we simply apply the triangular waveform instead of the tunneling bias and make use of the capacitance between tip and sample (typically 0.3--1\,pF) to examine the frequency response of the amplifier (and adjust it if required, e.g., after modifications of the cabling, changing the input capacitance). An example of such square wave output is presented in section \ref{sec:performance}. The triangle generator is also helpful for troubleshooting, e.g.\ in case a bad contact or a shorted bias voltage is suspected: One can quickly check the correct operation of the system without disconnecting the preamplifier.

\section{\label{sec:design}Amplifier Design}

Figure \ref{fig4_circuit}(a) shows the details of our amplifier circuit.  The operational amplifier of Fig.\ \ref{fig1_basics} is replaced by three stages, with gains of 10 and 5 for the 2nd and 3rd stage, respectively. Since the feedback resistor is connected to the output of the 3rd stage, the 2nd and 3rd stage do not increase the overall transimpedance, only the open-loop gain of the circuit. We use a 1\,\si{\giga\ohm} feedback resistor and a nominal output voltage range of $\pm 50$\,V for the last stage, which provides a current range of $\pm 50$\,nA. Together with the low thermal noise of the feedback resistor and the low input bias current of the first-stage amplifier (AD8615, 0.2\,pA typ.), this ensures a very large dynamic range as desired for a general-purpose STM preamplifier (see introduction). This dynamic range is achieved without the need of switching the feedback resistor, which would be difficult especially when putting the first stage into vacuum.

The first stage (OP1 and $R_\mathrm{f}$) can be placed in vacuum if desired. The AD8615 works well at liquid-nitrogen (LN2) temperatures. For operation at liquid-helium temperatures slight counterheating by its power dissipation is needed, thus it should be mounted with high thermal resistance to the cryostat. \cite{huber_2017}\,
\footnote{So far, we have operated a total of six AD8615 and AD8616 (the twin version of the AD8615) chips at cryogenic temperatures without problems. As these chips have been also used at cryogenic temperatures in other labs,\cite{huber_2017} they seem to work reliably at these temperatures.}
In our experience, standard RuO$_2$-based \si{\giga\ohm} resistors have no substantial temperature coefficient down to LN2 temperature; at LHe temperature the resistance increases (in our case, by $\approx 70$\%, requiring readjustment of the frequency compensation and different gain settings in the STM controller). One can use shielded cables for connecting to the first stage (to R$_\mathrm{f}$ and from OP1 output); we have not noticed any influence of 120\,pF capacitance from these lines to ground. We found that the cable shields must not be used as the only ground connections, however: Due to their inductance with respect to ground and the current due to the capacitance between shield and inner conductor, a voltage drop would occur along the cable shields. A ``solid'' ground connection via the cooling tubes of the cryostat or other massive parts in UHV is required to ensure stable operation of the amplifier.
If the first stage is outside vacuum, next to the rest of the circuit, it is important to take into account that the combined gain-bandwidth product of the first three stages is 1.2\,GHz (24\,MHz for the AD8615 in the first stage, gain 50 for 2nd and 3rd stage). Together with the high input impedance, this makes it clear that any stray capacitance between the input and the later stages must be intercepted by an electrostatic shield.

Independent of whether the first stage is in UHV or outside vacuum, it is important to avoid a ground plane below the feedback resistor. The capacitance between the resistive track and ground would lead to a large negative phase shift (delay) of the feedback signal, which is difficult to compensate and likely to cause oscillations. A further important consideration for the first stage is reducing the noise of its supply voltage. We found that the noise levels of standard voltage regulators lead to increased noise of the amplifier; we therefore use voltage dividers and large electrolytic capacitors (1000\,\si{\micro\farad}, with 470\,nF ceramic multi-layer capacitors in parallel) to provide a $+2.5$/$-1.5$\,V supply voltage to the AD8615 and at the same time suppress high-frequency noise. Trying to add inductors (33\,mH) for even better filtering did not yield any improvement of the noise performance.

The second stage amplifies the nominal $\pm 1$\,V output range of OP1 output to $\pm 10$\,V. It uses a fast operational amplifier, either OP37 (gain bandwidth product 63 MHz) or THS4631 (210\,MHz). With the latter also higher gains would be possible without introducing any sizable phase shift, i.e., without reducing the overall phase margin (for a gain of 10, the OP37 has the advantage of better DC accuracy due to a lower voltage offset). This stage also includes a voltage limiter, to avoid saturation of OP2 or OP3. The third stage, with $\pm 60$\,V supply voltage, provides the output voltage of nominally $\pm 50$\,V, to obtain a large dynamic range. The LTC6090-5 used for this stage offers a good compromise between speed (gain-bandwidth product 24\,MHz, slew rate 37\,\si{\volt/\micro\second} typ.) and voltage range (supply max.\ $\pm 70$\,V). Since the output of the amplifier (to the STM controller) should be $\pm 10$\,V, not $\pm 50$\,V, we use the output of the second stage as the $\pm 10$\,V amplifier output.

As mentioned in the introduction, a flat frequency response is obtained using a compensation network based on Fig.\ \ref{fig1_basics}(c). The potentiometer in series to C$_\mathrm{c}$ can be used to reduce the bandwidth if desired (reducing the overall noise); especially for fast STMs this is not needed because of the integrating (I) controller usually employed for constant-current STM imaging, which suppresses high-frequency noise. The resistor in series to the potentiometer should be chosen such that strong overshoot or oscillations cannot occur at the minimum setting of the bandwidth potentiometer; for an input capacitance of 35\,pF we found 470\,\si{\ohm} a suitable value.
The additional RC series circuit labelled ``mid-f compensation'' can be used for slightly tweaking the frequency response in the region between the $R_\mathrm{f}C_\mathrm{f}$ pole (1--2\,kHz) and the bandwidth limit. Suitable values for the mid-f-compensation components depend on details of the environment of the feedback resistor; for different geometries (breadboard, printed circuit board) we have found capacitors in the 100\,pF--1\,nF range useful (for values near the lower end, the trim potentiometer in series should have a higher value than shown, 200--500\,\si{\kilo\ohm}).

Setting the compensation network to obtain a flat frequency response ensures also a low phase shift of the feedback network and, hence, stability in the linear regime. Nevertheless, the circuit may oscillate when it enters the nonlinear regime; this is caused by additional delays in the feedback loop when an operational amplifier recovers from saturation or reaches its slew rate limit. Thus, without any additional measures, large signals or spikes at the input (which often occur in STM applications) could drive the circuit into oscillations. To suppress saturation of the 2nd and 3rd stages, the second stage includes a voltage limiter. As soon as the second stage reaches an output voltage slightly above 11\,V (either polarity), one of the two transistors (acting as common-base amplifier) will start conducting and provide an additional negative-feedback path strongly reducing the gain of the 2nd stage. The two antiparallel diodes suppress the influence of  transistor leakage currents or parasitic capacitances on the input of OP2; these currents instead find their way via a 10\,\si{\kilo\ohm} resistor to ground. Oscillations in the nonlinear regime can also occur without saturation of OP2 and OP3: The time delay of recovery from saturation of OP1 and the time for full output swing of OP3 limited by its slew rate are comparable; this situation is similar to two stages introducing a phase shift of \ang{-90} each, which will render the circuit unstable. This can be avoided by limiting the slew rate of OP3 for large voltage excursions. This is done by a high-pass circuit with a voltage divider at the output: If the output swing is large with a high slew rate, the output voltage of the voltage divider will be high enough to overcome the forward threshold of one of the two diodes. This additional negative-feedback path will reduce the slew rate. For ``normal'' signals, including high-frequency noise, this limiter remains inactive and does not introduce nonlinearity. 

\section{\label{sec:design}Further Considerations for an STM Preamplifier}

As for all sensitive measurements, where the signal source (here, the STM) and the signal processing (STM control electronics) are separated by some distance and/or have separate ground connections, it is essential to avoid ground loops. For STMs in ultrahigh vacuum (UHV), this cannot be done by running the transimpedance amplifier at the ground voltage of the STM control electronics, since the unavoidable capacitance from the current input to ground of the UHV chamber would then cause capacitive pickup of the difference of the ground potentials. This means that the transimpedance amplifier must use the ground of the UHV chamber and the STM controller should have a differential input for the output of the transimpedance amplifier. If this is not the case, the output of the transimpedance amplifier should be fed into an instrumentation amplifier (INA) taking its reference (output ground) from the STM controller. Essentially the same is true for the bias voltage, which must be supplied with respect to the ground potential of the transimpedance amplifier, i.e., with respect to UHV ground. For this purpose, it is best to have an instrumentation amplifier as a driver for the bias voltage (we use an AD8421, which is placed on the circuit board of the transimpedance amplifier). In addition, to avoid ground loops, the transimpedance amplifier (and the INA for driving the bias) should have its own power supply, with the ground connected to the UHV chamber, not to mains ground.

The usual bias voltage range of an STM is $\pm 10$\,V, which is higher than the permissible supply voltage of the first-stage OP1 (6\,V between the positive and negative supply rails for the AD8615). Since a short between the tip and sample of an STM cannot be ruled out, it is therefore necessary to limit the output current of the bias driver to the maximum current of the protection diodes of OP1 (5\,mA for the AD8615).

\begin{figure}
\includegraphics[width=8cm]{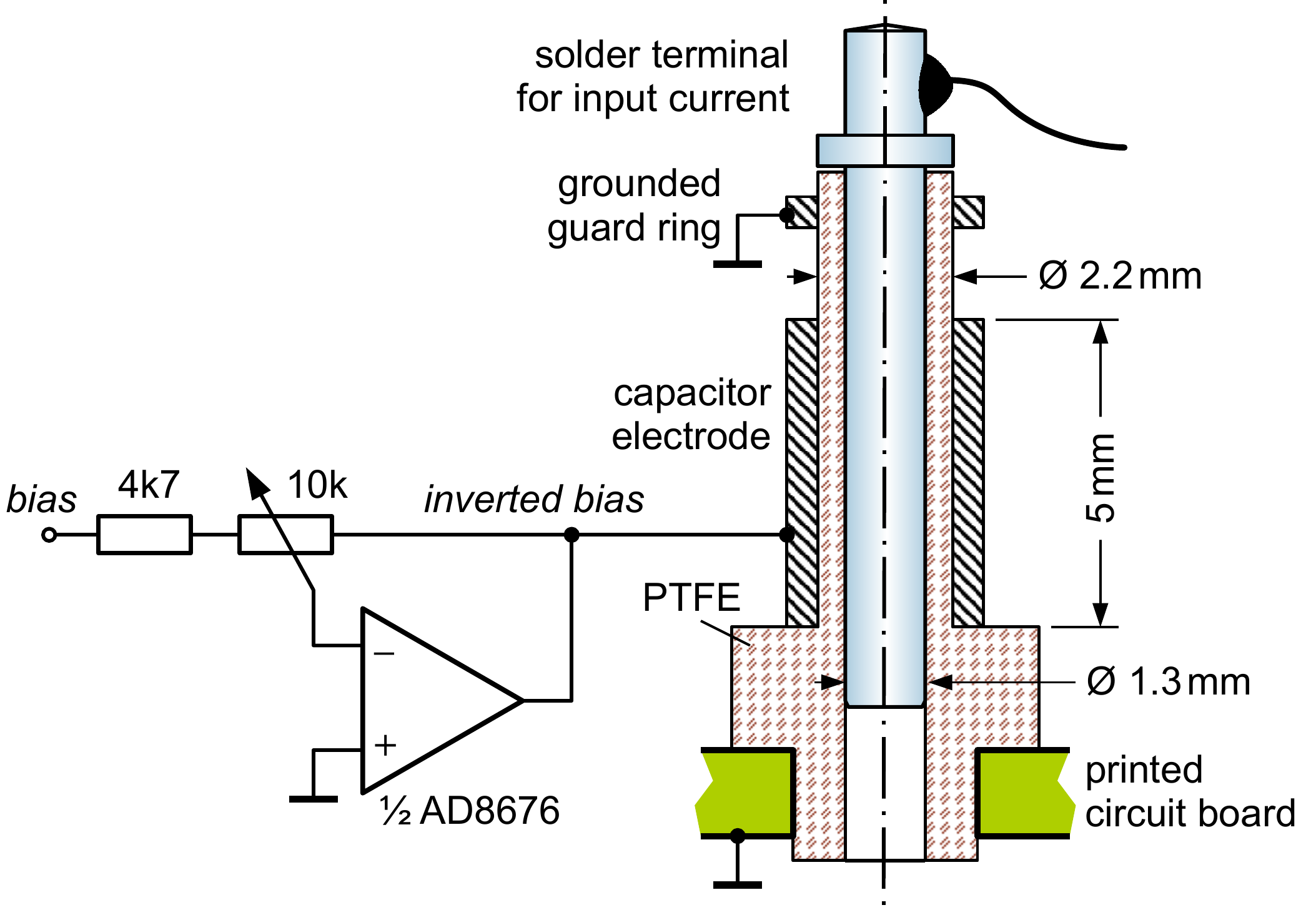}
\caption{\label{fig5_c_comp}Circuit for capacitance compensation and physical realization of a 1\,pF capacitor built into the solder terminal for the amplifier input.}
\end{figure}

For scanning tunneling spectroscopy (STS) with a lock-in amplifier, the choice of a good modulation frequency is difficult: On the one hand, it is desirable to use a high frequency, to escape the $1/f$ noise of the tunneling junction and have short settling times. On the other hand, at high frequencies the capacitance between tip and sample will lead to a large capacitive current, which can be much larger than the actual (modulated) tunneling current. This will make lock-in measurements very sensitive to the phase and also increase shot noise. Another problem caused by the tip-sample capacitance is the occurrence of current spikes when changing the bias voltage. The spikes have a large bandwidth and therefore disturb lock in-amplifiers, increasing the settling time required before a measurement can be taken. The spikes can be also detrimental for normal STM operation in constant-current mode, as lowering the magnitude of the bias will lead to a spike with a polarity opposite to that of the tunneling current, which can cause the STM controller to push the tip into the surface.
All these problems can be avoided by compensation of the capacitive current at the input of the transimpedance amplifier, as shown in Fig.\ \ref{fig5_c_comp}: We feed the inverted bias signal (with variable gain) into a capacitor with a similar capacitance as the tip-sample capacitance.  As the bias can be a few volts, it is important to avoid any leakage current either through this capacitor or at its surface, since even a \si{\tera\ohm} leakage will result in an input current of a few pA. For the amplifier with the first stage outside vacuum, we therefore use a homemade cylindrical capacitor built into the PTFE-supported input terminal of the amplifier, with grounded guard electrodes between the input and the inverted bias (Fig.\ \ref{fig5_c_comp}). While this compensation works well, we do not have sufficient experience with STS measurements to judge whether it can come close to the performance of highly optimized radio-frequency methods.\cite{bastiaans_amplifier_2018}

\section{\label{sec:performance}Performance of the Amplifier}

\begin{figure}
\includegraphics[width=8cm]{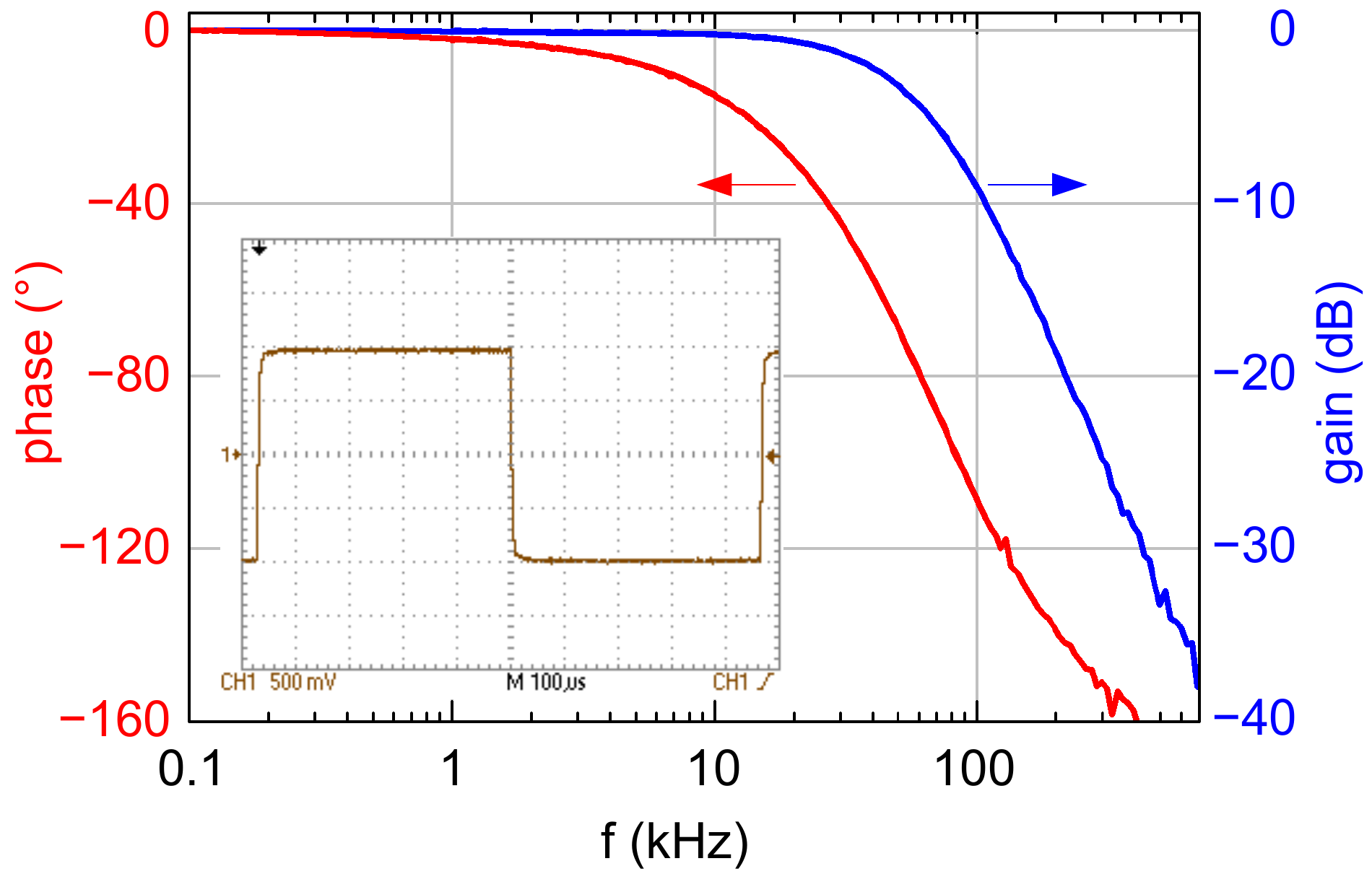}
\caption{\label{fig6_expGain}Frequency response of the circuit in Fig.\ \ref{fig4_circuit} with 35\,pF input capacitance. The data were measured with an AC voltage applied to the tip-sample capacitance and division by $j\omega$ as explained in section \ref{sec:measurement}. 0\,dB corresponds to the DC transimpedance of the circuit (1\,\si{\giga\ohm} at the output of the 3rd stage, 1/5\,\si{\giga\ohm} at the $\pm 10$\,V output of stage 2). The inset shows the oscilloscope trace with a 1.07\,kHz triangle wave connected to the sample, using the capacitive coupling to the STM tip to operate the amplifier as a differentiator and check the frequency response (The oscilloscope was in $\times$8 averaging mode for noise reduction).}
\end{figure}

Depending on the exact layout of the environment of the feedback resistor, the bandwidth of the amplifier in Fig.\ \ref{fig4_circuit} is about 100\,kHz with zero input capacitance and 50\,kHz with 35\,pF input capacitance (Fig.\ \ref{fig6_expGain}). These values can be increased by increasing the gain of the second stage (we could obtain up to 200\,kHz at an input capacitance of $\approx 2$\,pF). Unfortunately, this makes accurate compensation of the frequency response more critical (to avoid oscillations). In principle, one could use bootstrapping to avoid the reduction of the bandwidth by the input capacitance.\cite{hoyle_shunt_1999} This would require the first stage to work with a fixed, positive gain, which is problematic due to the proximity of the noninverting input pin and the negative supply of most operational amplifiers, as mentioned in the introduction. Another possibility to increase the bandwidth is compensation at the output as shown in Fig.\ \ref{fig1_basics}(b); for the curve in Fig.\ \ref{fig6_expGain}, this could easily increase the bandwidth to 125\,kHz. For our application, a bandwidth of 50\,kHz is sufficient, however.

\begin{figure}
\includegraphics[width=8.5cm]{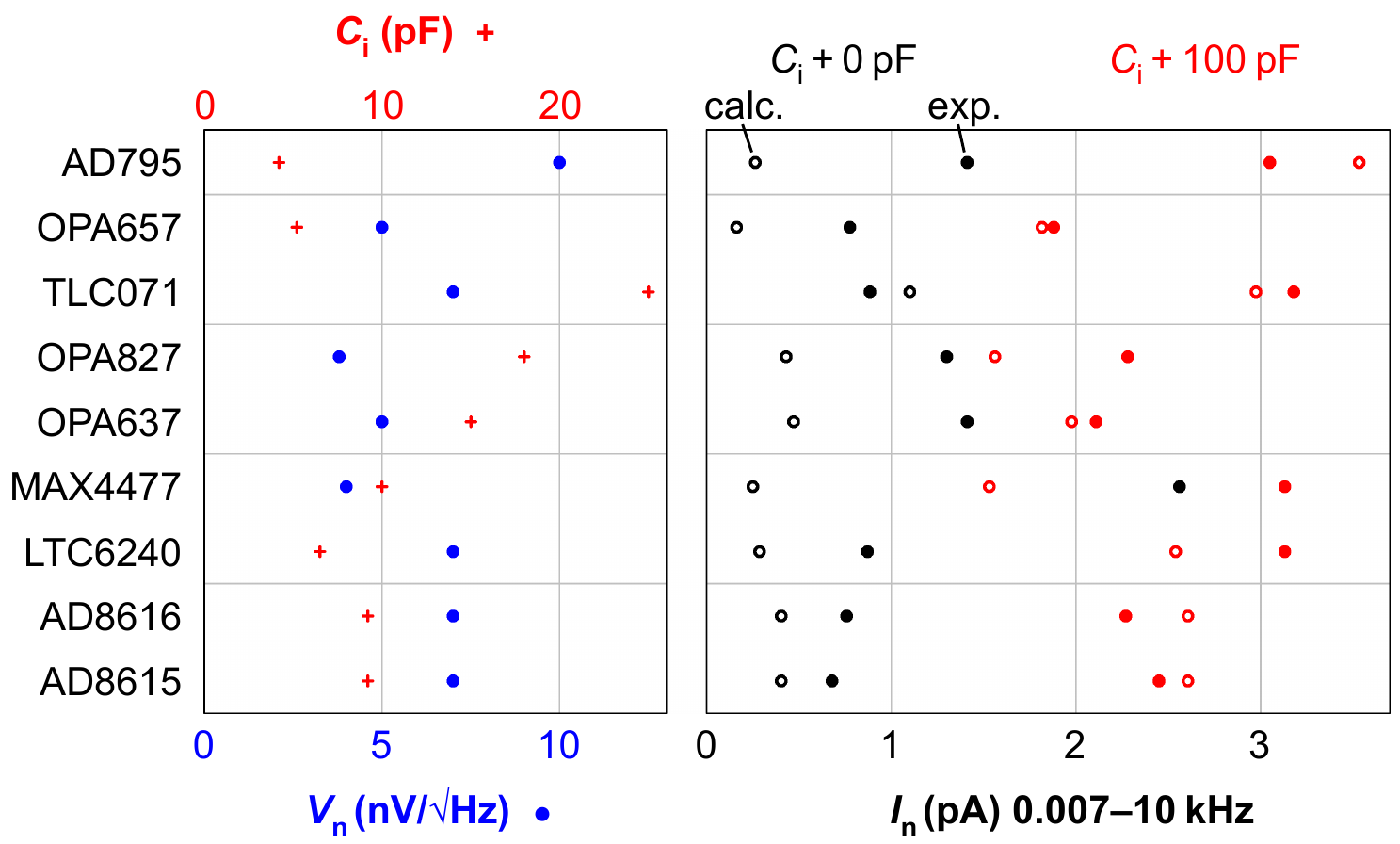}
\caption{\label{fig7_opampNoiseData}Comparison of different operational amplifiers. The left panel shows the input voltage noise at 10\,kHz (filled blue circles) and the input capacitance (red crosses), both according to the datasheets. The right panel shows the total noise measured with a 7\,Hz high-pass filter and a 10\,kHz 4th-order low-pass filter (filled circles) as well as the noise calculated for this frequency range (open circles), with and without an additional input capacitance of 100\,pF (red and black, respectively). Except for the AD8615/8616, where very similar results were found for several chips, only one specimen was tested for each operational amplifier type.}
\end{figure}

Concerning the noise performance, we have tried our circuit with different operational amplifier types, with and without an additional 100\,pF capacitance at the input, to simulate the amplifier being connected to the STM by long cable (as for cryogenic STMs with the amplifier outside vacuum) or very close to the STM (cable capacitance negligible). We have measured the bandwidth-integrated noise with a 1st-order high-pass filter (7\,Hz, to suppress the DC component) and a 10\,kHz 4th-order low-pass filter. For comparison, we have also calculated the noise as expected from the Johnson noise of the feedback resistor, as well as values from the datasheet for the input voltage noise $v_\mathrm{n}$ and input capacitance of each operational amplifier type, using eq. (\ref{eq:noiseCin}). The noise powers of these two contributions have been multiplied by the frequency response of the low-pass filter and integrated over the frequency.
Figure \ref{fig7_opampNoiseData} shows these values, together with the values for the input voltage noise and input capacitance used for the calculation
\footnote{We did not take the frequency dependence of the input voltage noise $v_\mathrm{n}$ into account but rather took the values at 10\,kHz, where the $v_\mathrm{n}$ contribution is largest. This is justified since for the operational amplifiers under consideration and the relevant frequency range, $v_\mathrm{n}$ only weakly depends on the frequency.}.
Especially for the data without additional input capacitance (black data points in Fig.\ \ref{fig7_opampNoiseData}), it is obvious that the calculations (open circles) strongly underestimate the noise in most cases; as explained in section \ref{sec:noise} we attribute this to capacitive coupling of voltage noise at the source terminals of the input FETs to their gate. According to eq. (\ref{eq:Ceff}), this additional noise contribution has similar consequences as an additional input capacitance, and its influence should vanish if a large external input capacitance is added. Indeed, this can be seen in Fig.\ \ref{fig7_opampNoiseData}: When adding 100\,pF input capacitance, the spread between the experimental and calculated noise values becomes smaller in most cases, and sometimes the experimental values are even better than the calculated ones (indicating that the particular operational amplifier used by us performs better than stated in the datasheet). The additional noise contribution according to eq.\ (\ref{eq:noiseUnbalanced2}) means that selecting operational amplifiers according to input voltage noise and input capacitance does not guarantee good noise performance in a transimpedance amplifier, as clearly seen in
Fig.\ \ref{fig7_opampNoiseData}. Among the amplifiers examined by us, at low additional input capacitance, the AD8615 and AD8616 (dual version of the AD8615) perform best, even if they do not offer exceptionally low input voltage noise according to the datasheet. This type also excels when used for pickup of tiny signals from a high-impedance quartz crystal in non-contact AFM \cite{huber_2017}. With 100\,pF additional input capacitance, reasoning according to eq.\ (\ref{eq:noiseCin}) holds and the best performance is delivered by operational amplifiers with very low input voltage noise.

\begin{figure}
\includegraphics[width=6.5cm]{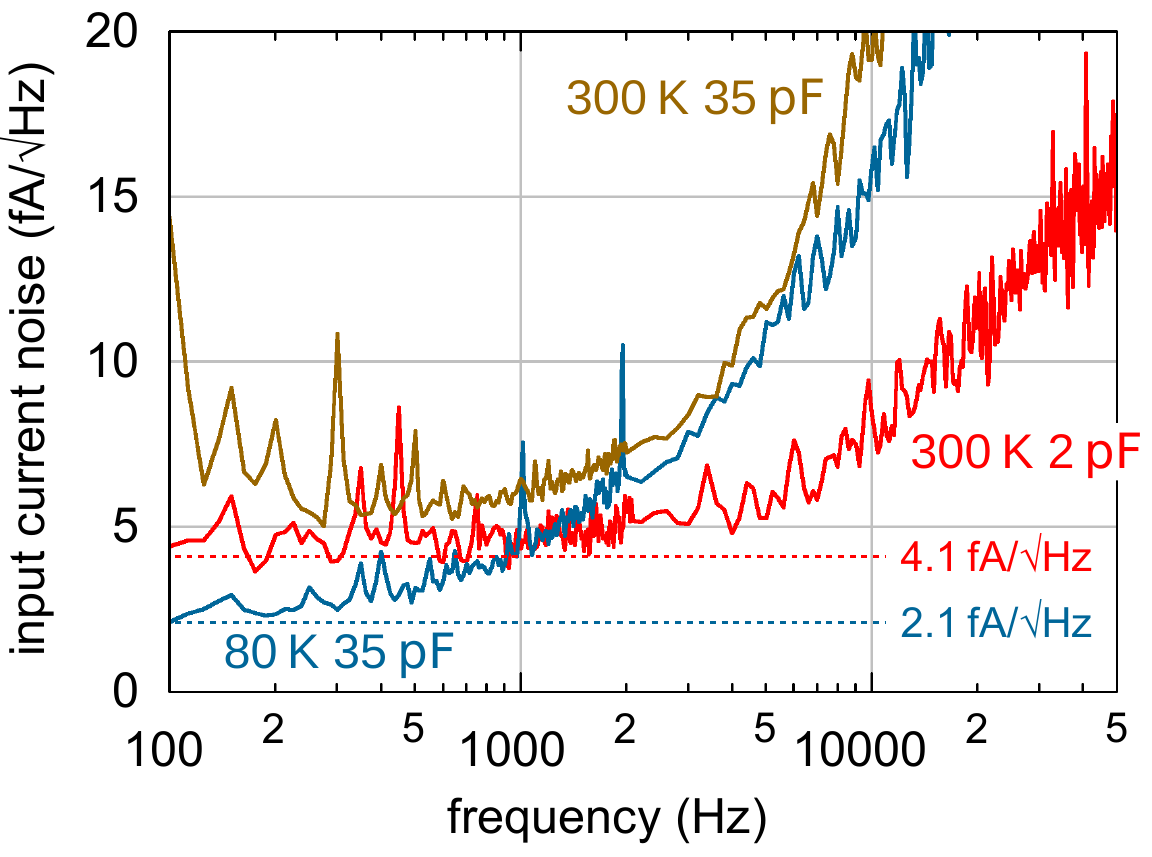}
\caption{\label{fig8_noiseSpectra}Noise spectra of two amplifiers according to Fig.\ \ref{fig4_circuit}, with the first stage at $T=80$\,K and $C_\mathrm{in} \approx 35$\,pF (cable capacitance), as well as an amplifier with all stages at room temperature and two different input capacitance values ($C_\mathrm{in}=2$ and 35\,pF, not including the $\approx 10$\,pF input capacitance of the AD8615 operational amplifier). The peaks below 500\,Hz are due to insufficient shielding (line frequency overtones; for the room-temperature amplifier at $C_\mathrm{in}=35$\,pF also including magnetic induction up to $\approx 1$\,kHz). The peaks at 1.0 and 2.0\,kHz in the spectrum of the low-temperature amplifier are attributed to electromagnetic interference from a turbomolecular pump.}
\end{figure}

Fig.\ \ref{fig8_noiseSpectra} shows exemplary noise spectra of our amplifier, with the first stage at room temperature (2 and 35\,pF input capacitance, red and brown respectively), as well as with the first stage in a liquid-nitrogen cooled STM head, with $\approx 35$\,pF capacitance of the cable to the tunneling junction (blue). It is obvious that the noise floor at low frequencies decreases as the feedback resistor is cooled (dotted lines mark the calculated Johnson noise), whereas the noise at higher frequencies strongly increases with increasing input capacitance. The extra noise of the room-temperature amplifier with $C_\mathrm{in}=35$\,pF at low frequencies ($\lesssim 1$\,kHz) is mainly at integer multiples of the line frequency and caused by inductive pick-up of magnetic stray fields. First experiments with the STM at $T=8$\,K (not shown) indicate a clearly reduced noise below 100\,Hz (to $\approx$ 1/2 the value at 80\,K) due to reduced Johnson noise, but higher noise above 1\,kHz (by $\approx 30\%$), possibly due to increased input voltage noise of the AD8615 or increased noise of the current source driving the input FETs (cf.\ Fig.\ \ref{fig3_inputNoise}).

\begin{figure*}
\includegraphics[width=17cm]{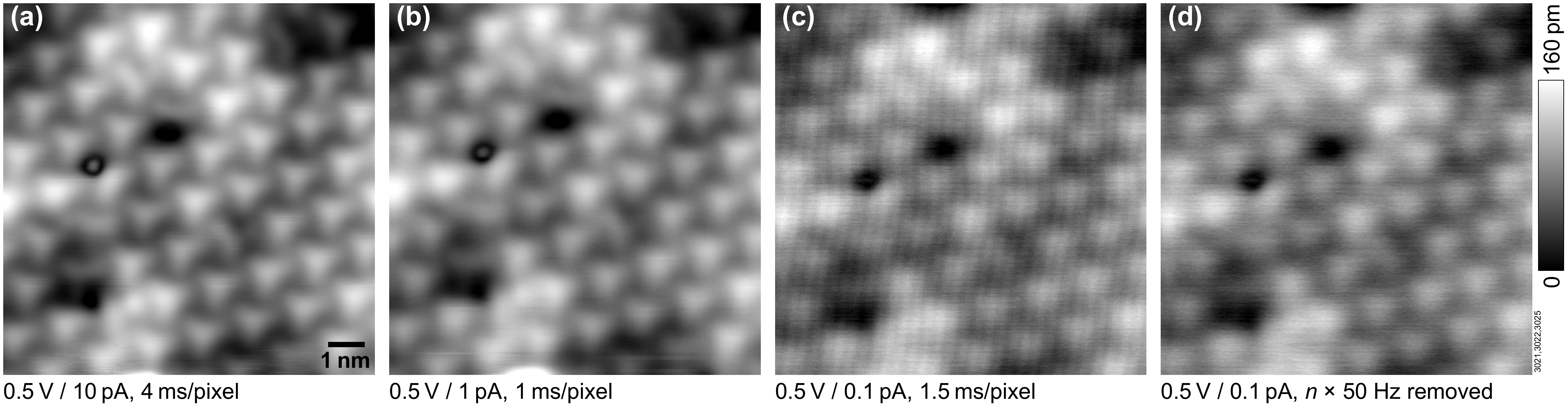}
\caption{\label{fig9_STM}Constant-current STM images of water dissociatively adsorbed at the the stoichiometric In$_2$O$_3$(111) surface\cite{wagner_In2O3+water_2017} acquired at $T=8$\,K, with setpoints (tunneling current) of (a) 10\,pA, (b) 1\,pA, and (c) 0.1\,pA. The data are unfiltered except for the subtraction of a planar background. Frame (d) shows the data of (c) with suppression of line-frequency overtones.}
\end{figure*}

An example of STM images acquired with our amplifier (again, with $\approx 35$\,pF cable capacitance) is shown in Fig.\ \ref{fig9_STM}.
\footnote{The STM images were acquired with a Tribus STM head from Sigma Surface Science (now Scienta Omicron) and an electrochemically etched W tip. Based on the amplitude of noise peaks at typical frequencies of mechanical vibrations, we estimate that mechanical vibrations above 2\,Hz are in the order of magnitude of 0.1\,pm, thus their contribution to the noise is negligible.}
With the first stage of the amplifier mounted to the STM operating at $\approx 8$\,K, reducing the tunneling current from (a) 10 to (b) 1\,pA does not lead to any obvious degradation of the image quality. An image acquired at 0.1\,pA shows line frequency interference (50\,Hz and overtones, $\Delta z_\mathrm{RMS}=1.6$\,pm, probably due to capacitive pick-up from cabling in the STM, e.g., piezo lines). After removal of the line-frequency overtones (d), the image clearly shows some noise, but the main surface features (protrusions indicating the hydroxyls at the surface\cite{wagner_In2O3+water_2017}) are still visible.

\begin{figure}
\includegraphics[width=8.5cm]{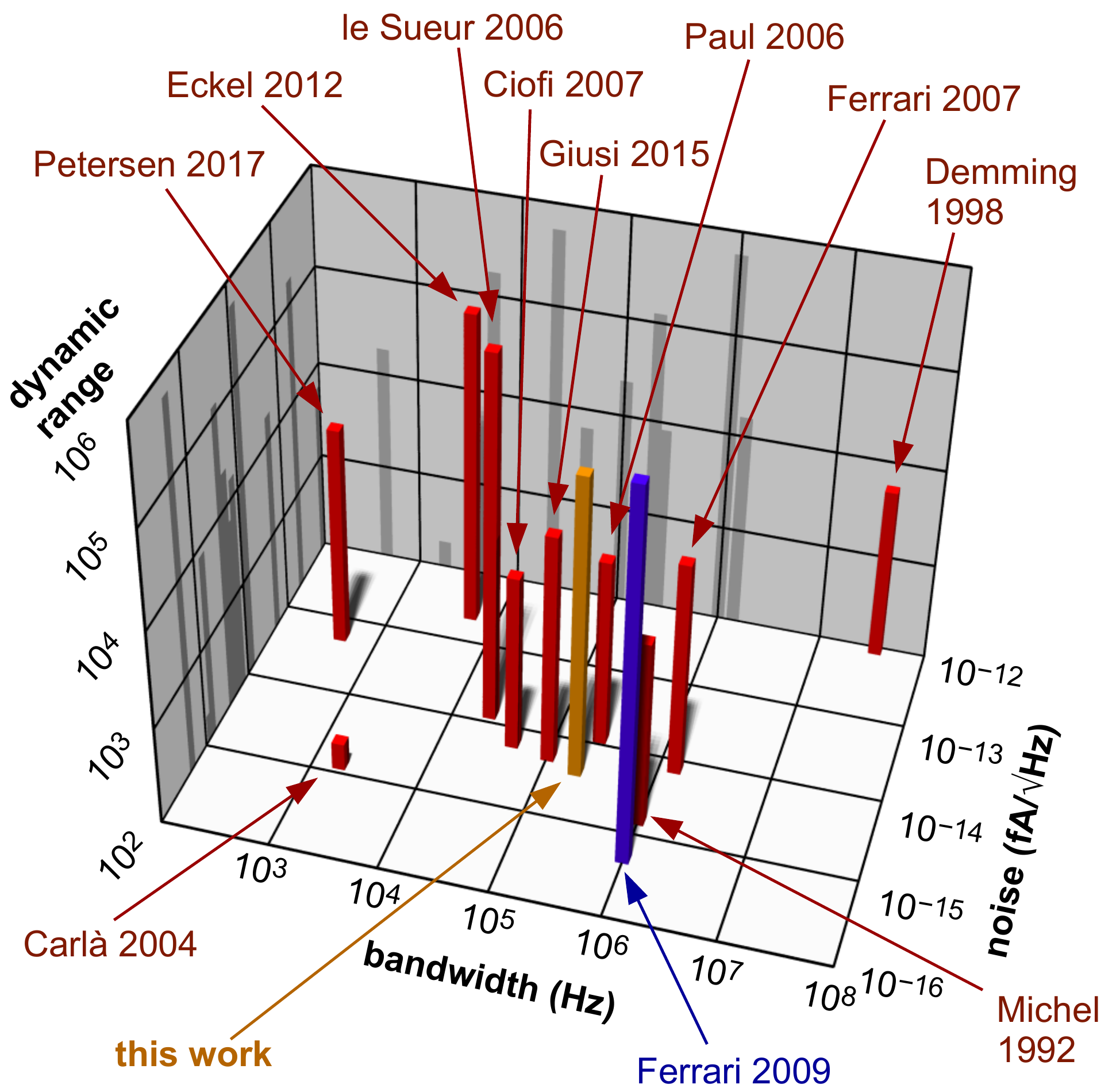}
\caption{\label{fig10_comparison}Comparison of transimpedance amplifiers described in the literature. Except for ``le Sueur 2006'', data are for small or zero input capacitance and room-temperature operation. The input current noise is given at 1\,kHz and the dynamic range is based on a lower limit given by a signal-to-noise ratio of 10\,dB at 1\,kHz bandwidth. The orange bar stands for the amplifier presented in this work (with a second-stage gain of $\approx 20$), blue for a specially developed integrated circuit; all others are red. Except for le Sueur 2006, the feedback resistor is at room temperature in all cases. The references are
Carl{\`a} 2004 \cite{carla_development_2004}, 
Ciofi 2007 \cite{ciofi_new_2007}, 
Demming 1998 \cite{demming_wide_1998},
Eckel 2012 \cite{eckel_note_2012} (three-range amplifier, data for range 2),
Ferrari 2007 \cite{ferrari_wide_2007},
Ferrari 2009 \cite{ferrari_ultra-low-noise_2009},
Giusi 2015 \cite{giusi_ultra-low-noise_2015},
le Sueur 2006 \cite{le_sueur_room-temperature_2006},
Michel 1992 \cite{michel_low-temperature_1992},
Paul 2006 \cite{paul_updated_2006},
Petersen 2017 \cite{petersen_circuit_2017}.
}
\end{figure}

Finally, Figure \ref{fig10_comparison} shows a comparison of the present work with several transimpedance amplifiers described in the literature. We show the maximum $-3$\,dB bandwidth achieved, the input current noise at 1\,kHz (when not given estimated from the Johnson noise of the feedback resistor) and the dynamic range. For the latter, we have assumed that the lower limit of the current range is given by a signal-to-noise ratio of 10\,dB at 1 kHz bandwidth, for simplicity assuming a frequency-independent noise density (equal to that at 1\,kHz).
The full-range current, when not given, was estimated from the circuit, assuming an output range of $\pm 12$\,V for operational amplifiers that can be operated with $\pm 15$\,V supplies. With one exception noted below, all data are for zero or very low input capacitance (data for input capacitance values typical for a cable between a UHV STM and an atmospheric-side amplifier were not available for most designs). In Fig.\ \ref{fig10_comparison}, better performance is indicated by a position of the bar further to the lower right and a taller bar. While there are a few amplifiers that are on the ``better'' side from ours, we note that all these have substantial limitations not present in our design. The amplifier labelled Ferrari 2009 \cite{ferrari_ultra-low-noise_2009} is a specially designed CMOS chip with nonlinear elements (FETs) replacing $R_\mathrm{f}$. It is uncertain whether these will cause crossover distortions and other problems due to nonlinearity; furthermore, without special precautions CMOS amplifiers tend to have rather high input voltage noise, which would cause additional noise according to eq.\ (\ref{eq:noiseCin}) as soon as any input capacitance is added. As mentioned in the introduction, the amplifier labelled Michel 1992 \cite{michel_low-temperature_1992} has the problem of large input voltage offset, and the Ferrari 2007 design \cite{ferrari_wide_2007} is in this form not directly usable for STM because it essentially consists of two separate amplifiers for the AC and DC components of the signal. The data labelled le Sueur 2006 \cite{le_sueur_room-temperature_2006} are not directly comparable with the others because they are for a room-temperature amplifier with the feedback resistor (14\,\si{\mega\ohm}) in a millikelvin cryostat, which strongly reduces the Johnson noise; on the other hand these data include the influence of a cable between the STM and amplifier.

\section{\label{sec:applications}Notes on Further Applications}

While our amplifier was developed for STMs, it is well-suited for many other applications. Its large bandwidth makes it useful for specimen current imaging \cite{farley1988} and electron-beam-induced conductivity (EBIC) \cite{leamy_charge_1982} in scanning electron microscopy (SEM). At sufficiently low data rates and high beam current, if the specimen noise is dominated by shot noise (c.f.\ Fig.\ \ref{fig2_noise}), we can argue that specimen current measurement can even lead to less noise than a conventional secondary electron (SE) detector (Everhart-Thornley detector): The specimen current is the difference between the incoming and the secondary electron current. If the SE yield is close to unity (at low beam voltages), this difference is small compared to the SE current, therefore its shot noise (which is proportional to the square root of the current) is also smaller. In that case, also any fluctuations of the primary beam current will affect the  specimen current less than the SE current. Specimen current imaging should be also of advantage for environmental SEM of low-Z materials, which have low backscattering yield and therefore low gas-ionisation signal.

For photodiode applications, where only one polarity is relevant, our amplifier could be modified using an asymmetric supply for OP3 (e.g., $-10$ and $+110$\,V), which would give it a dynamic range up to 100\,nA with a 1\,\si{\giga\ohm} resistor. This would require either a gain of 10 for the 3rd stage or an asymmetric supply, e.g., $-25$ and $+5$\,V for the 2nd stage (since OP2 is substantially faster than OP3, this would be preferable to keep phase shifts low). Also the transistor-base voltages of the limiter in the 2nd stage must be adapted for the full-scale voltage of the 2nd stage. It should be noted that the idea of using higher-than-usual voltages for the output stage of photodiode transimpedance amplifiers has been proposed earlier, but that previous circuit has only provided a bandwidth of 28\,kHz with a 10\,\si{\mega\ohm} feedback resistor and 3\,pF photodiode capacitance \cite{brisebois_photodiode_2007}; our circuit offers a larger bandwidth and ten times lower Johnson noise (which is especially  relevant at low frequencies and low input currents).

\section{\label{sec:conclusions}Conclusions}

We have presented a transimpedance amplifier designed for scanning tunneling microscopy. The amplifier is based on commercially available standard operational amplifiers and provides a larger bandwidth than most previous designs, combined with very low noise. In addition, our amplifier provides an exceptionally large dynamic range without range switching. The first stage of the amplifier, including the feedback resistor, can be placed in UHV and also works at cryogenic temperatures, which makes it well-suited for low-temperature STM. We have described a fast and simple way to judge the frequency response of a transimpedance amplifier, and we have provided evidence for a source of noise that was previously not discussed in the design of transimpedance amplifiers and is caused by a similar mechanism as the induced gate noise of FETs. We have also discussed various additional considerations for the use of the amplifier for STM applications. Although the amplifier presented was designed for STMs it may find many other applications.

\begin{acknowledgments}
This work was supported by the Austrian Science Fund (FWF) projects F4505 (Functional Oxide Surfaces and Interfaces ``FOSXI'') and Wittgenstein Prize Z 250.

The data that support the findings of this study are available from the corresponding author upon reasonable request.
\end{acknowledgments}

\bibliography{STM_preamp}

\end{document}